\documentclass[11pt,preprintnumbers,titlepage]{revtex4-2}%
\usepackage[latin9]{inputenc}
\usepackage{amsmath}
\usepackage{amssymb}
\usepackage{graphicx}
\usepackage[unicode=true,  bookmarks=false,  breaklinks=false,pdfborder={0 0
1},backref=section,colorlinks=false]{hyperref}
\usepackage{amsfonts}
\usepackage{subfigure}
\usepackage{cleveref}
\usepackage{listings}
\usepackage{xcolor}
\usepackage{array}
\usepackage{url}
\usepackage{color}%
\hypersetup{
 colorlinks,linkcolor=blue,citecolor=blue,urlcolor=blue}
\makeatletter
\@ifundefined{textcolor}{}{
\definecolor{BLACK}{gray}{0}
\definecolor{WHITE}{gray}{1}
\definecolor{RED}{rgb}{1,0,0}
\definecolor{GREEN}{rgb}{0,1,0}
\definecolor{BLUE}{rgb}{0,0,1}
\definecolor{CYAN}{cmyk}{1,0,0,0}
\definecolor{MAGENTA}{cmyk}{0,1,0,0}
\definecolor{YELLOW}{cmyk}{0,0,1,0}
}

\makeatother
\begin{document}
\preprint{CTP-SCU/2021038}
\title{Quasinormal Modes of Black Holes with Multiple Photon Spheres}
\author{Guangzhou Guo$^{a}$}
\email{gzguo@stu.scu.edu.cn}
\author{Peng Wang$^{a}$}
\email{pengw@scu.edu.cn}
\author{Houwen Wu$^{a,b}$}
\email{hw598@damtp.cam.ac.uk}
\author{Haitang Yang$^{a}$}
\email{hyanga@scu.edu.cn}
\affiliation{$^{a}$Center for Theoretical Physics, College of Physics, Sichuan University,
Chengdu, 610064, China}
\affiliation{$^{b}$Department of Applied Mathematics and Theoretical Physics, University of
Cambridge, Wilberforce Road, Cambridge, CB3 0WA, UK}

\begin{abstract}
For a static and spherically symmetric black hole, a photon sphere is composed
of circular null geodesics of fixed radius, and plays an important role in
observing the black hole. Recently, in an Einstein-Maxwell-scalar model with a
non-minimal coupling between the scalar and electromagnetic fields, a class of
hairy black holes has been found to possess two unstable and one stable
circular null geodesics on the equatorial plane, corresponding to three photon
spheres outside the event horizon. In this paper, we study quasinormal modes
of the scalar field, which are associated with these circular null geodesics,
in the hairy black hole spacetime. In the eikonal regime with $l\gg1$, the
real part of the quasinormal modes is determined by the angular velocity of
the corresponding circular geodesics. The imaginary part of the quasinormal
modes associated with the unstable circular null geodesics encodes the
information about the Lyapunov exponent of the corresponding circular
geodesics. Interestingly, we find long-lived and sub-long-lived modes, which
are associated with the stable and one of the unstable circular null
geodesics, respectively. Due to tunneling through potential barriers, the
damping times of the long-lived and sub-long-lived modes can be exponentially
and logarithmically large in terms of $l$, respectively.

\end{abstract}
\maketitle
\tableofcontents

{}

\section{Introduction}

In the last decade, gravitational waves from a binary black hole merger were
successfully detected by LIGO and Virgo \cite{Abbott:2016blz}, and
subsequently the first image of a supermassive black hole at the center of
galaxy M87 was photographed by the Event Horizon Telescope (EHT)
\cite{Akiyama:2019cqa,Akiyama:2019brx,Akiyama:2019sww,Akiyama:2019bqs,Akiyama:2019fyp,Akiyama:2019eap}%
, which opens a new era of black hole physics. Due to the event horizon, a
black hole is a dissipative system, eigenmodes of which are quasinormal modes.
The final stage of a binary black hole merger is ringdown, in which the
gravitational waveforms are described by a superposition of quasinormal modes
\cite{Berti:2007dg}. Quasinormal modes with complex frequencies have vast
applications in black hole physics
\cite{Ferrari:1984zz,Nollert:1999ji,Horowitz:1999jd,Berti:2003ud,Myung:2008pr,Berti:2009kk,Konoplya:2011qq,Yang:2012he,Cook:2016fge,Myung:2018jvi,Konoplya:2019hlu,Blazquez-Salcedo:2020jee}%
. In particular, a linear perturbation can induce a discrete set of
quasinormal modes, whose imaginary part is related to the damping time scale.
The spectrum of quasinormal modes therefore provides a perspective on the
study of stability of the background spacetime
\cite{Doneva:2010ke,Blazquez-Salcedo:2018jnn,Myung:2018vug,Myung:2019oua,Zou:2020zxq,Guo:2021zed}%
. Moreover, the most dominant quasinormal modes can be used to check the
validity of the strong cosmic censorship conjecture
\cite{Cardoso:2017soq,Gan:2019jac,Gan:2019ibg}.

Since exotic horizonless objects, e.g., wormholes and ultra-compact objects
(UCOs), have optical observations similar to black holes, they have recently
attracted great attention
\cite{Lemos:2008cv,Cunha:2017wao,Cunha:2018acu,Shaikh:2018oul,Huang:2019arj,Wielgus:2020uqz,Yang:2021diz,Peng:2021osd}%
. Interestingly, a reflecting boundary in the wormhole or UCO spacetime can
produce a set of time-delay echoes, which are characterized by quasinormal
modes \cite{Mark:2017dnq,Bueno:2017hyj,Cardoso:2019rvt,Ou:2021efv}.
Furthermore, UCOs have been conjectured to suffer from instabilities due to
the existence of a family of long-lived quasinormal modes, which appear in the
neighborhood of a stable circular null geodesic
\cite{Cardoso:2014sna,Keir:2014oka,Guo:2021bcw}. For instance, a linear
ergoregion instability associated with long-lived modes may occur for a
spinning object with a sufficiently high rotation speed
\cite{Friedman1978ErgosphereI,Chirenti:2008pf,Pani:2008bzt}.

Intriguingly, unstable null geodesics have been revealed to be closely related
to a class of quasinormal modes of perturbations in the black hole spacetime
\cite{Ferrari:1984zz,Nollert:1999ji,Cardoso:2008bp,Yang:2012he,Konoplya:2017wot,Jusufi:2019ltj,Cuadros-Melgar:2020kqn,Qian:2021aju}%
. In \cite{Ferrari:1984zz}, null geodesics were first found to be connected
with quasinormal modes in Schwarzschild and slowly rotating Kerr black holes.
Using the WKB approximation in the eikonal limit, the authors of
\cite{Cardoso:2008bp} elaborated the connection in the static, spherically
symmetric and asymptotically flat black hole background. To be more specific,
it verified that the real part of quasinormal modes is proportional to the
angular velocity of the corresponding unstable circular null geodesic, while
the imaginary part is determined by the Lyapunov exponent of the orbit.
Furthermore, the relation between null geodesics and quasinormal modes was
generalized to Kerr black holes of arbitrary spin in \cite{Yang:2012he}, which
showed an extra precession modification in the real part compared to
non-rotating black holes.

The No-hair theorem asserts that a black hole is uniquely determined by its
three parameters, i.e., mass, electric charge and angular momentum
\cite{Israel:1967wq,Carter:1971zc,Ruffini:1971bza}. However, hairy black holes
with extra freedom have been constructed in various models, which provide
counter-examples to the no-hair theorem
\cite{Volkov:1989fi,Bizon:1990sr,Greene:1992fw,Luckock:1986tr,Droz:1991cx,Kanti:1995vq,Mahapatra:2020wym}%
. Recently, a type of Einstein-Maxwell-scalar (EMS) models with a non-minimal
coupling between the scalar and electromagnetic fields have been extensively
studied in the literature
\cite{Herdeiro:2018wub,Myung:2018vug,Astefanesei:2019pfq,Fernandes:2019kmh,Fernandes:2019rez,Peng:2019cmm,Zou:2019bpt,Astefanesei:2020qxk,Blazquez-Salcedo:2020nhs,Fernandes:2020gay,Guo:2020zqm,Myung:2020dqt,Myung:2020ctt,Wang:2020ohb,Gan:2021pwu,Gan:2021xdl,Guo:2021zed,Guo:2021ere}%
. In the EMS models, the non-minimal coupling destabilizes scalar-free black
holes and induce the onset of spontaneous scalarization to form hairy black
holes with a scalar hair \cite{Herdeiro:2018wub}. In
\cite{Myung:2019oua,Zou:2020zxq,Myung:2020etf}, the stability of the hairy
black holes was analyzed by calculating their quasinormal modes of various
perturbations. Decaying quasinormal modes may suggest that the hairy black
holes are the endpoints of the dynamic evolution from unstable scalar-free
black hole solutions.

Surprisingly, it showed that in a certain parameter regime of the hairy black
holes, there exist two unstable and one stable null circular null geodesics on
the equatorial plane, which indicates three photon spheres of different sizes
outside the event horizon \cite{Gan:2021pwu,Gan:2021xdl}. Due to a double-peak
structure appearing in the potential of the photon radial motion, the
existence of two unstable photon spheres can remarkably affect the optical
appearance of black holes illuminated by the surrounding accretion disk, e.g.,
leading to bright rings of different radii in the black hole image
\cite{Gan:2021pwu} and significantly increasing the flux of the observed image
\cite{Gan:2021xdl}. The relation between null geodesics and quasinormal modes
has been rarely reported for black holes with more than one photon sphere.
Moreover, multiple photon spheres appearing in some spacetime signal the
existence of long-live modes, which may render the spacetime unstable
\cite{Cardoso:2014sna,Keir:2014oka,Guo:2021bcw}. Therefore, it is of great
interest to study quasinormal modes of the hairy black holes endowed with
three photon spheres. Note that multiple photon spheres have recently been
reported in different black hole models
\cite{Liu:2019rib,Brihaye:2021mqk,Huang:2021qwe}.

In this paper, we use the WKB method to calculate quasinormal modes localized
at circular null geodesics of the hairy black holes with three photon spheres.
The rest of the paper is organized as follows. In Section
\ref{Circular Orbits in the EMS Model}, we study null circular geodesics of
hairy black holes in the EMS model, as well as the orbital stability by
evaluating the Lyapunov exponent. Subsequently, quasinormal modes trapped at
different circular null geodesics are obtained in Section
\ref{sec:QNMs at CRs}. We conclude our main results in Section \ref{Sec:Conc}.
The Appendix \ref{sec:appd} is devoted to derivations of some WKB formulas. We
set $16\pi G=1$ throughout this paper.

\section{Hairy Black Holes}

\label{Circular Orbits in the EMS Model}

In this section, we first briefly review spherically symmetric hairy black
hole solutions in the EMS model. Subsequently, we study circular geodesics for
photons around the hairy black holes and compute the corresponding Lyapunov exponents.

\subsection{Black Hole Solution}

In the EMS model, the action is given by
\begin{equation}
S=\int d^{4}x\sqrt{-g}\left[  R-2\partial_{\mu}\phi\partial^{\mu}%
\phi-e^{\alpha\phi^{2}}F^{\mu\nu}F_{\mu\nu}\right]  , \label{eq:Action}%
\end{equation}
where the scalar field $\phi$ is minimally coupled to the metric field and
non-minimally coupled to the electromagnetic field $A_{\mu}$. Here, $F_{\mu
\nu}=\partial_{\mu}A_{\nu}-\partial_{\nu}A_{\mu}$ is the electromagnetic field
strength tensor, and $e^{\alpha\phi^{2}}$ is the coupling function between
$\phi$ and $A_{\mu}$. Following \cite{Herdeiro:2018wub,Guo:2021zed}, we
restrict our attention to static, spherically symmetric and asymptotically
flat black hole solutions with the generic ansatz
\begin{align}
ds^{2}  &  =-N\left(  r\right)  e^{-2\delta\left(  r\right)  }dt^{2}+\frac
{1}{N\left(  r\right)  }dr^{2}+r^{2}\left(  d\theta^{2}+\sin^{2}\theta
d\varphi^{2}\right)  ,\nonumber\\
A_{\mu}dx^{\mu}  &  =V\left(  r\right)  dt\text{ and }\phi=\phi\left(
r\right)  . \label{eq:ansatz}%
\end{align}
The equations of motion are then given by
\begin{align}
N^{\prime}\left(  r\right)   &  =\frac{1-N\left(  r\right)  }{r}-\frac{Q^{2}%
}{r^{3}e^{\alpha\phi^{2}\left(  r\right)  }}-rN\left(  r\right)  \phi
^{\prime2}\left(  r\right)  ,\nonumber\\
\left[  r^{2}N\left(  r\right)  \phi^{\prime}\left(  r\right)  \right]
^{\prime}  &  =-\frac{\alpha\phi\left(  r\right)  Q^{2}}{e^{\alpha\phi
^{2}\left(  r\right)  }r^{2}}-r^{3}N\left(  r\right)  \phi^{\prime3}\left(
r\right)  ,\nonumber\\
\delta^{\prime}\left(  r\right)   &  =-r\phi^{\prime2}\left(  r\right)
,\label{eq:NLEqs}\\
V^{\prime}\left(  r\right)   &  =\frac{Q}{r^{2}e^{\alpha\phi^{2}\left(
r\right)  }}e^{-\delta\left(  r\right)  },\nonumber
\end{align}
where primes denote derivatives with respect to $r$, and the integration
constant $Q$ is interpreted as the electric charge of the black hole solution.

To find black holes solutions from the non-linear ordinary differential
equations $\left(  \ref{eq:NLEqs}\right)  $, one needs to impose proper
boundary conditions at the event horizon $r_{h}$ and the spatial infinity,
\begin{align}
N(r_{h}) &  =0\text{, }\delta(r_{h})=\delta_{0}\text{, }\phi(r_{h})=\phi
_{0}\text{, }V(r_{h})=0\text{,}\nonumber\\
N(\infty) &  =1\text{, }\delta(\infty)=0\text{, }\phi(\infty)=0\text{,
}V(\infty)=\Phi\text{,}\label{eq:BC}%
\end{align}
where $\Phi$ is the electrostatic potential. The two parameters $\delta_{0}$
and $\phi_{0}$ determine the asymptotic behavior of the solutions in the
vicinity of the horizon. Moreover, the black hole mass $M$, which is related
to the ADM mass, can be obtained via $M=\lim\limits_{r\rightarrow\infty
}r\left[  1-N(r)\right]  /2$. In this paper, we set $M=1$ and use a shooting
method to numerically solve eqn. $\left(  \ref{eq:NLEqs}\right)  $ for black
hole solutions matching the boundary conditions $\left(  \ref{eq:BC}\right)
$. It is manifest that the scalar-free solutions with $\phi=0$ (i.e.,
Reissner-Nordstr\"{o}m black holes) can exist in the EMS model. Nevertheless,
we focus on hairy black holes with the non-trivial profile of the scalar field
$\phi$. For instance, we exhibit the profile of the metric functions for the
hairy black hole solution with $\alpha=0.9$ and $Q=1.066$ in the left panel of
FIG. \ref{Hbh-plot}.

\subsection{Circular Null Geodesics}

Owing to strong gravity near a black hole, photons are forced to travel in
circular null geodesics on photon spheres, which play an important role in
determining properties of the black hole image seen by a distant observer
(e.g., the size of the black hole shadow). Here, circular null geodesics of
the spherically symmetric hairy black hole are studied. Without loss of
generality, we consider a photon moving on the equatorial plane with
$\theta=\pi/2$. To obtain equatorial geodesics, we start from the Lagrangian
\begin{equation}
\mathcal{L}=\frac{1}{2}\left(  -N\left(  r\right)  e^{-2\delta\left(
r\right)  }\dot{t}^{2}+\frac{1}{N\left(  r\right)  }\dot{r}^{2}+r^{2}%
\dot{\varphi}^{2}\right)  , \label{eq:nullLag}%
\end{equation}
where dots denote derivatives with respect to the affine parameter $\tau$. The
generalized canonical momenta for this Lagrangian are defined as
\begin{align}
-p_{t}  &  =N\left(  r\right)  e^{-2\delta\left(  r\right)  }\dot
{t}=E,\nonumber\\
p_{\varphi}  &  =r^{2}\dot{\varphi}=L,\label{eq:CM}\\
p_{r}  &  =\frac{1}{N}\dot{r}.\nonumber
\end{align}
Note that the metric of the hairy black hole spacetime is independent of $t$
and $\varphi$. So the spacetime admits two Killing vectors, which are
associated with the conserved energy $E$ and momentum $L$, respectively, in
eqn. $\left(  \ref{eq:CM}\right)  $. Varying the Lagrangian $\left(
\ref{eq:nullLag}\right)  $ with respect to $r$ yields the radial equation of
motion for the photon,
\begin{equation}
\frac{d}{d\tau}\frac{\partial\mathcal{L}}{\partial\dot{r}}=\frac
{\partial\mathcal{L}}{\partial r}. \label{eq:radial Eq}%
\end{equation}
With the help of eqn. $\left(  \ref{eq:CM}\right)  $, eqn. $\left(
\ref{eq:radial Eq}\right)  $ becomes
\begin{equation}
e^{-2\delta\left(  r\right)  }\dot{r}^{2}=E^{2}-\frac{e^{-2\delta\left(
r\right)  }N\left(  r\right)  }{r^{2}}L^{2}, \label{eq:radial eq}%
\end{equation}
which describes a null geodesic. For later use, one can introduce the
geometric potential as
\begin{equation}
V_{\text{geo}}\left(  r\right)  =\frac{e^{-2\delta\left(  r\right)  }N\left(
r\right)  }{r^{2}}. \label{eq: Veff}%
\end{equation}
Accordingly, a null circular geodesic at $r=r_{c}$ can appear, provided that
the conditions $V_{\text{geo}}\left(  r_{c}\right)  =E^{2}/L^{2}$ and
$V_{\text{geo}}^{\prime}\left(  r_{c}\right)  =0$ are satisfied.

\begin{figure}[ptb]
\begin{centering}
\includegraphics[scale=0.85]{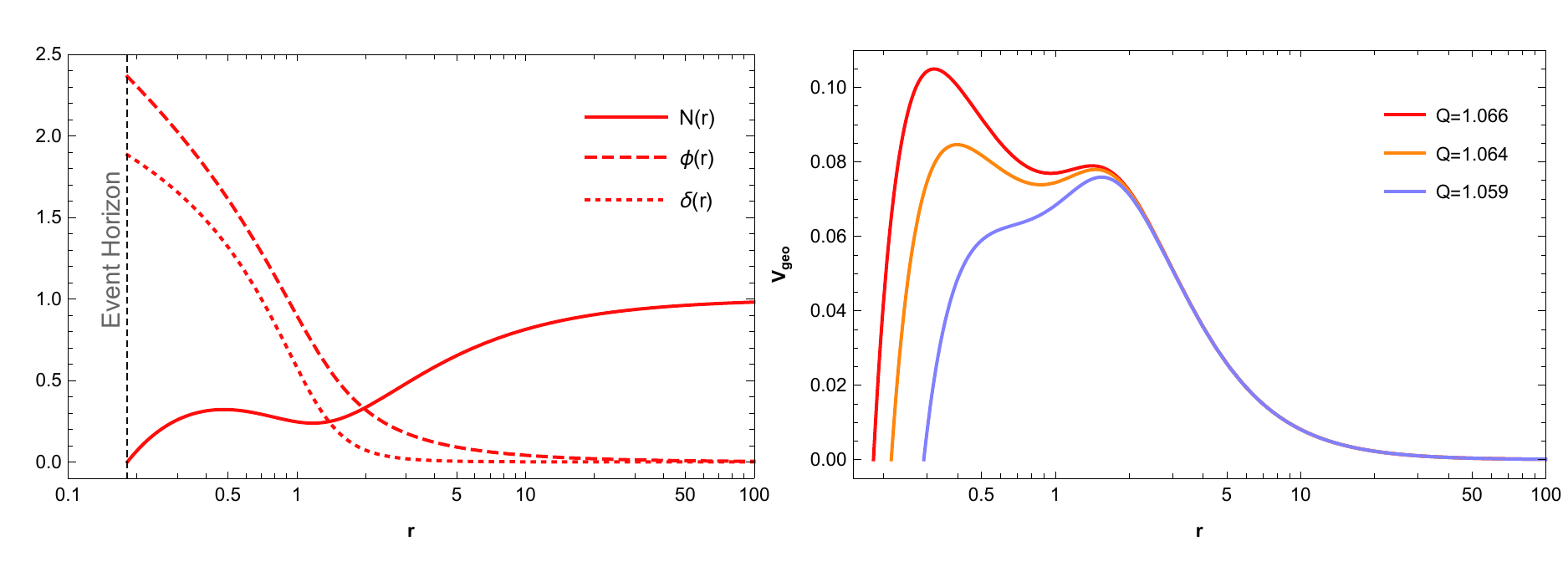}
\par\end{centering}
\caption{Metric functions and geometric potential for hairy black hole
solutions, where we take $\alpha=0.9$. \textbf{Left}: The metric functions
$N\left(  r\right)  $ (solid line), $\phi\left(  r\right)  $ (dashed line) and
$\delta\left(  r\right)  $ (dotted line) for the hairy black hole solution
with $Q=1.066$ are plotted outside the event horizon (vertical dashed line).
\textbf{Right}: The geometric potential for different hairy black holes with
$Q=1.066$ (red line), $Q=1.064$ (orange line) and $Q=1.059$ (blue line). For a
large value of charge (e.g., the red and orange lines), the geometric
potential possesses a double-peak structure with one local minimum and two
local maxima. By contrast, the double-structure disappears as the charge
decreases (e.g., the blue line), leaving only a single maximum of the
geometric potential.}%
\label{Hbh-plot}%
\end{figure}

In FIG. \ref{Hbh-plot}, we present the geometric potential for hairy black
holes with $\alpha=0.9$ for $Q=1.066,\text{ }1.064\text{ and }1.059$ in the
right panel. When $Q=1.059$ (the blue line), the geometric potential possesses
a single maximum outside the event horizon. Intriguingly for $Q=1.066$ and
$1.064$ (the red and orange lines), it displays a double-peak structure with
one minimum and two maxima, which implies that there exist three null circular
geodesics located at the extrema.

\subsection{Lyapunov Exponent}

The Lyapunov exponent is proposed to characterize the rate of separation of
adjacent trajectories in the phase space \cite{Cornish:2003ig,Cardoso:2008bp}.
In a dynamical system, the sign of the Lyapunov exponent can be used to
determine whether adjacent trajectories converge or not. Specifically,
positive Lyapunov exponents correspond to the divergent trajectories, while
negative ones to the convergent trajectories. Therefore, we can study the
stability of null circular orbits around a hairy black hole by evaluating
their Lyapunov exponents. Moreover, the Lyapunov exponent can be closely
related to quasinormal modes of black holes. Indeed, it showed in
\cite{Cardoso:2008bp} that, if the geometric potential has a single maximum,
the imaginary part of quasinormal modes is determined by the Lyapunov exponent
of the unstable circular orbit at the maximum in the eikonal regime.

To describe circular orbits in spherically symmetric spacetime, we focus on a
two dimensional phase space spanned by $X_{i}\left(  t\right)  \equiv\left(
p_{r},r\right)  $. The equations of motion in this phase space can be
schematically written as
\begin{equation}
\frac{dX_{i}}{dt}=H_{i}\left(  X_{j}\right)  . \label{eq:PhaseEOM}%
\end{equation}
To obtain the Lyapunov exponent of a given orbit, we need to linearize eqn.
$\left(  \ref{eq:PhaseEOM}\right)  $ around the orbit,
\begin{equation}
\frac{d\delta X_{i}\left(  t\right)  }{dt}=K_{ij}\left(  t\right)  \delta
X_{j}\left(  t\right)  , \label{eq:Leq}%
\end{equation}
where the linear stability matrix $K_{ij}\left(  t\right)  $ is
\begin{equation}
K_{ij}\left(  t\right)  =\left.  \frac{\partial H_{i}}{\partial X_{j}%
}\right\vert _{X_{i}\left(  t\right)  }=\left(
\begin{array}
[c]{cc}%
0 & d\left(  \dot{t}^{-1}\dot{p_{r}}\right)  /dr\\
\dot{t}^{-1}N & 0
\end{array}
\right)  . \label{eq:SMat}%
\end{equation}
The solution to the linearized equation $\left(  \ref{eq:Leq}\right)  $ can be
expressed as
\begin{equation}
\delta X_{i}\left(  t\right)  =L_{ij}\left(  t\right)  \delta X_{j}\left(
0\right)  ,
\end{equation}
where the evolution matrix $L_{ij}\left(  t\right)  $ satisfies
\begin{equation}
\frac{dL_{ij}\left(  t\right)  }{dt}=K_{im}\left(  t\right)  L_{mj}\left(
t\right)  ,
\end{equation}
and $L_{ij}\left(  0\right)  =\delta_{ij}$. The principal Lyapunov exponent is
related to the determination of the eigenvalues of $L_{ij}$, i.e.,
\begin{equation}
\lambda=\lim_{t\rightarrow\infty}\frac{1}{t}\log\frac{L_{jj}\left(  t\right)
}{L_{jj}\left(  0\right)  }.
\end{equation}
From eqn. $\left(  \ref{eq:SMat}\right)  $, the principal Lyapunov exponent
can be written as
\begin{equation}
\lambda=\pm\sqrt{\dot{t}^{-1}N\frac{d}{dr}\left(  \dot{t}^{-1}\dot{p_{r}%
}\right)  },
\end{equation}
where we choose the $+$ sign for the Lyapunov exponent \cite{Cardoso:2008bp}.
Specifically, the Lyapunov exponent of a circular orbit at $r=r_{c}$ can be
expressed in terms of the geometric potential,
\begin{equation}
\left.  \lambda\right\vert _{r=r_{c}}=\left.  \sqrt{-\frac{L^{2}e^{2\delta}%
}{2\dot{t}^{2}}V_{\text{geo}}^{\prime\prime}}\right\vert _{r=r_{c}}=\left.
\sqrt{-\frac{1}{2V_{\text{geo}}}\frac{d^{2}}{dx^{2}}V_{\text{geo}}}\right\vert
_{r=r_{c}}, \label{eq:Lya}%
\end{equation}
where the tortoise coordinate $x$ is defined by $dx/dr\equiv e^{\delta\left(
r\right)  }N^{-1}\left(  r\right)  $. For a circular null orbit located at a
local maximum of the geometric potential with $V_{\text{geo}}^{\prime\prime
}<0$, the Lyapunov exponent $\lambda$ is positive, which implies that the
orbit is unstable under small perturbations. On the other hand, a circular
orbit located at a local minimum has $V_{\text{geo}}^{\prime\prime}<0$ and
hence has a\ purely imaginary value of the Lyapunov exponent, which implies
that the orbit is stable. Therefore, a black hole with a double-peak geometric
potential (e.g., $Q=1.066$ and $1.064$ in FIG. \ref{Hbh-plot}) has two
unstable circular null geodesics at the local maxima and a stable one at the
local minimum. In addition, we introduce the angular velocity $\Omega$ of the
circular orbit at $r=r_{c}$ via
\begin{equation}
\left.  \Omega\right\vert _{r=r_{c}}=\left.  \frac{\dot{\varphi}}{\dot{t}%
}\right\vert _{r=r_{c}}=\left.  \sqrt{V_{\text{geo}}}\right\vert _{r=r_{c}},
\label{eq:angular velocity}%
\end{equation}
which can be related to the real part of quasinormal modes
\cite{Cardoso:2008bp}.

\section{Quasinormal Modes}

\label{sec:QNMs at CRs}

In this section, we consider the perturbation of the scalar field and compute
its quasinormal modes that are associated with the circular null geodesics in
the hairy black hole background $\left(  \ref{eq:ansatz}\right)  $. For a
scalar perturbation of mode
\begin{equation}
\delta\phi=e^{-i\omega t}\frac{\Psi_{\omega l}\left(  r\right)  Y_{lm}\left(
\theta,\phi\right)  }{r},
\end{equation}
we can separate the angular variables and express the linearized equation for
$\Psi_{\omega l}\left(  r\right)  $ in the following general form,
\begin{equation}
\left(  \frac{d^{2}}{dx^{2}}+\omega^{2}-V_{l}\left(  r\right)  \right)
\Psi_{\omega l}\left(  r\right)  =0, \label{eq: Pert eq}%
\end{equation}
where the effective potential is
\begin{equation}
V_{l}\left(  r\right)  =\frac{e^{-2\delta\left(  r\right)  }N\left(  r\right)
}{r^{2}}\left[  l\left(  l+1\right)  +1-N\left(  r\right)  -\frac{Q^{2}}%
{r^{2}e^{\alpha\phi\left(  r\right)  ^{2}}}-\left(  \alpha+2\alpha^{2}%
\phi\left(  r\right)  ^{2}\right)  \frac{Q^{2}}{r^{2}e^{\alpha\phi\left(
r\right)  ^{2}}}\right]  .
\end{equation}
In the eikonal regime $\left(  l\gg1\right)  $, the effective potential
reduces to
\begin{equation}
V_{l}\left(  r\right)  \simeq l^{2}\frac{e^{-2\delta\left(  r\right)
}N\left(  r\right)  }{r^{2}}=l^{2}V_{\text{geo}}\left(  r\right)  ,
\label{eq:effgeoPotential}%
\end{equation}
where $V_{\text{geo}}\left(  r\right)  $ is the aforementioned geometric
potential. It is worth emphasizing that the full perturbed fields around the
hairy black hole background were considered in \cite{Myung:2019oua}. In the
eikonal limit, the scalar perturbation decouples with other perturbations, and
the scalar perturbative equation then reduces to eqn. $\left(
\ref{eq: Pert eq}\right)  $ with the effective potential $\left(
\ref{eq:effgeoPotential}\right)  $.

\begin{figure}[ptb]
\begin{centering}
\includegraphics{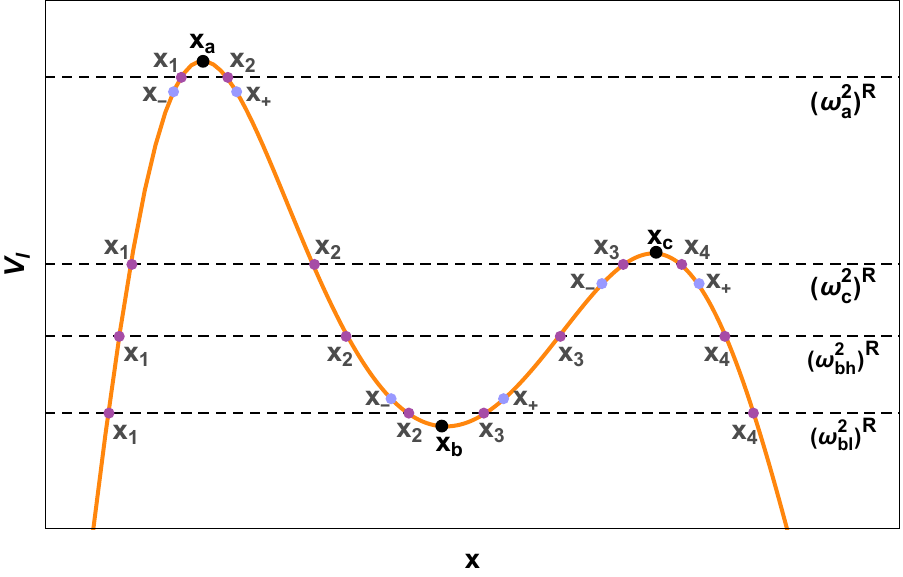}
\par\end{centering}
\caption{The effective potential $V_{l}$ as a function of the tortoise
coordinate $x$. Here we present a local region of the effective potential for
the hairy black hole with $\alpha=0.9$ and $Q=1.066$. Note that the effective
potential tends to vanish as $x\rightarrow\pm\infty$. The horizontal dashed
lines represent the real parts of quasinormal modes $\omega^{2}$ in different
cases, which intersect the effective potential at the turning points labelled
by $x_{1}$, $x_{2}$, $x_{3}$ and $x_{4}$. The black points denote the extrema
of $V_{l}$ at $x_{a}$, $x_{b}$ and $x_{c}$, respectively, which correspond to
the circular null geodesics in the eikonal limit. Around the extrema, the
effective potential can be approximated by a parabola between the blue points
$x_{-}$ and $x_{+}$.}%
\label{Veffx}%
\end{figure}

In FIG. \ref{Veffx}, we plot the effective potential $V_{l}$ with $l=3$ for
the hairy black hole solution with $\alpha=0.9$ and $Q=1.066$. Like the
geometric potential, $V_{l}$ displays a double-peak structure, i.e., two local
maxima and one minimum. The effective potential $V_{l}$ with $l>3$ is found to
have a similar profile. Since quasinormal modes are determined by the
effective potential $V_{l}$, multiple local extrema of $V_{l}$ could lead to
distinct types of quasinormal modes. In the rest of the section, we apply the
WKB approximation to studying quasinormal modes trapped at the extrema of the
effective potential with a double-peak structure. These modes can be
interpreted as photons moving in the circular null geodesics determined by
$V_{\text{geo}}\left(  r\right)  $, which leak out slowly. Since we are
interested in the large $l$ limit, the WKB approximation can be accurate
enough for computing quasinormal modes. For comparison, we perform a direct
integration to numerically solve eqn. $\left(  \ref{eq: Pert eq}\right)  $ to
obtain the exact quasinormal modes of interest for the hairy black hole with
$\alpha=0.9$ and $Q=1.066$. Note that quasinormal modes occur only when
appropriate boundary conditions are imposed, i.e., purely ingoing modes at
$x=-\infty$ (at the event horizon) and purely outgoing modes at $x=+\infty$
(at the spatial infinity). Without causing any ambiguity, we denote the
geometric potential, the effective potential and the metric functions with
respect to the tortoise coordinate $x$ by $V_{\text{geo}}\left(  x\right)  $,
$V_{l}\left(  x\right)  $, $N\left(  x\right)  $ and $\delta\left(  x\right)
$, respectively.

\subsection{Modes at Global Maximum}

We first discuss the quasinormal modes of frequency $\omega_{a}$ trapped at
the global maximum of the effective potential at $x=x_{a}$. For such modes,
the $\left(  \omega_{a}^{2}\right)  ^{R}$ line lies close to the global
maximum and intersects the potential at the turning points $x_{1}$ and $x_{2}%
$, around which the WKB approximation fails. There exists some small positive
constant $\delta$ such that the WKB approximation is valid for $x<x_{1}%
-\delta$ and $x>x_{2}+\delta$. Therefore, the solution of eqn. $\left(
\ref{eq: Pert eq}\right)  $ can be approximated by the WKB expansions in
$\left(  -\infty,x_{1}-\delta\right)  $ and $\left(  x_{2}+\delta
,+\infty\right)  $, which match the ingoing and outgoing boundary conditions,
respectively. In the vicinity of $x=x_{a}$, the potential can be approximated
by a parabola. If $\left(  \omega_{a}^{2}\right)  ^{R}$ is close enough to the
global maximum, there exist $x_{-}$ and $x_{+}$ with $x_{-}<x_{1}-\delta
<x_{2}+\delta<$ $x_{+}$, such that the effective potential is well
approximated by a parabola for $x_{-}<x<x_{+}$ (see FIG. \ref{Veffx}). In
$\left(  x_{-},x_{+}\right)  $, eqn. $\left(  \ref{eq: Pert eq}\right)  $ with
an approximated parabolic potential can be exactly solved in terms of
parabolic cylinder functions. Furthermore, a complete solution requires that
the two WKB expansions should be smoothly connected by the exact solution in
the overlapping regions $\left(  x_{-},x_{1}-\delta\right)  $ and $\left(
x_{2}+\delta,x_{+}\right)  $. The matching procedure then gives
\cite{Schutz:1985km}
\begin{equation}
\frac{\omega_{a}^{2}-V_{l}\left(  x_{a}\right)  }{\sqrt{-2V_{l}^{(2)}\left(
x_{a}\right)  }}=i\left(  n+\frac{1}{2}\right)  \text{ with }n=0,1,2\cdots,
\label{eq:QNMcond1}%
\end{equation}
where the superscript of $V_{l}^{(2)}\left(  x_{a}\right)  $ denotes the
second derivative over the tortoise coordinate $x$.

The matching condition $\left(  \ref{eq:QNMcond1}\right)  $ leads to a set of
discrete quasinormal modes $\omega_{a}$, labelled by the integer $n$. In the
eikonal regime $\left(  l\gg1\right)  $, the quasinormal modes reduce to
\begin{equation}
\omega_{a}=l\sqrt{V_{\text{geo}}\left(  x_{a}\right)  }-i\left(  n+\frac{1}%
{2}\right)  \sqrt{-\frac{V_{\text{geo}}^{(2)}\left(  x_{a}\right)
}{2V_{\text{geo}}\left(  x_{a}\right)  }}, \label{eq:QNM1}%
\end{equation}
where $x_{a}$ becomes the global maximum of the geometric potential in the
eikonal limit. Using eqns. $\left(  \ref{eq:Lya}\right)  $ and $\left(
\ref{eq:angular velocity}\right)  $, one can express the quasinormal modes
$\left(  \ref{eq:QNM1}\right)  $ as
\begin{equation}
\omega_{a}=\Omega_{a}l-i\left(  n+\frac{1}{2}\right)  \lambda_{a},
\label{eq:gmQC}%
\end{equation}
where $\Omega_{a}$ and $\lambda_{a}$ are the angular velocity and the Lyapunov
exponent of the unstable circular orbit at the global maximum of the geometric
potential, respectively. Interestingly, since the Lyapunov exponent of the
unstable circular orbit describes the instability timescale of the geodesic
motion, the Lyapunov exponent contributes to the imaginary part of the
quasinormal modes. Note that the case with a single maximum of the geometric
potential was found to have the same relation between quasinormal modes and
circular null geodesics as eqn. $\left(  \ref{eq:gmQC}\right)  $
\cite{Cardoso:2008bp}.

\subsection{Long-lived Modes at Minimum}

In this subsection, the quasinormal modes of low and high excitations trapped
at the minimum of the effective potential are derived via the WKB
approximation, respectively. We find that there exist long-lived quasinormal
modes, which is related to the stable circular null geodesic with a purely
imaginary Lyapunov exponent. It is noteworthy that the existence of long-lived
modes has been reported in the spacetime of ECOs with an unstable photon orbit
\cite{Cardoso:2014sna}.

\begin{figure}[ptb]
\begin{centering}
\includegraphics[scale=0.6]{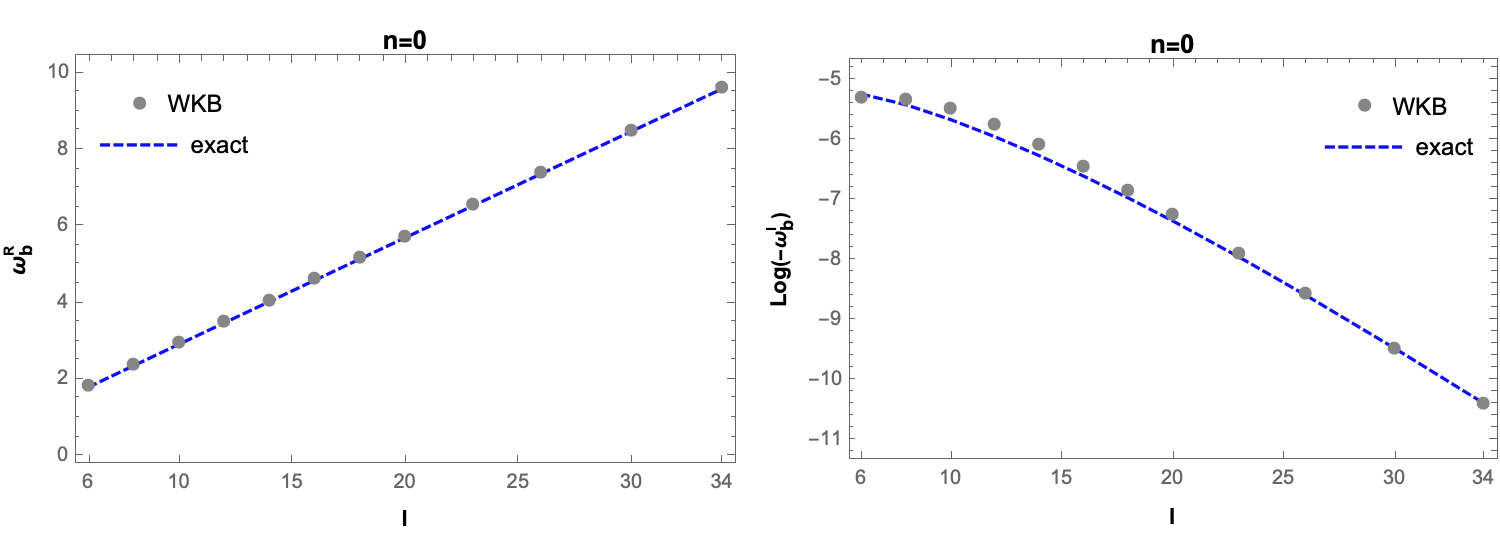}
\par\end{centering}
\caption{ Real (\textbf{Left}) and imaginary (\textbf{Right}) parts of the
lowest-lying quasinormal modes ($n=0$) trapped at the stable circular null
geodesic. Blue dashed lines represent the exact modes obtained by a numerical
method of direct integration, while gray dots denote the WKB results. The real
part $\omega_{b}^{R}$ increases monotonically with $l$. For a large $l$, the
ratio between $\omega_{b}^{R}$ and $l$ is the angular velocity of the circular
geodesic. The magnitude of $\omega_{b}^{I}$ decreases exponentially as $l$
increases, which indicates a class of long-lived modes.}%
\label{Long-lived modes}%
\end{figure}

We first discuss the low-lying quasinormal modes of frequency $\omega_{bl}$.
As illustrated in FIG. \ref{Veffx}, the $\left(  \omega_{bl}^{2}\right)  ^{R}$
line intersects the effective potential at the turning points $x_{1}$, $x_{2}%
$, $x_{3}$ and $x_{4}$. Suppose the WKB approximation is valid at a distance
$\delta$ away from these turning points, where $\delta>0$ is a small constant.
If the $\left(  \omega_{bl}^{2}\right)  ^{R}$ line is close enough to the
local minimum at $x=x_{b}$, both $x_{2}-\delta$ and $x_{3}+\delta$ can lie in
the interval $\left(  x_{-},x_{+}\right)  $, in which the effective potential
is well approximated by a parabolic expansion. With the approximated parabolic
potential, the exact solution of eqn. $\left(  \ref{eq: Pert eq}\right)  $ can
be expressed in terms of parabolic cylinder functions for $x\in\left(
x_{-},x_{+}\right)  $. In addition, the effective potential can be
approximated by a linear function near $x_{1}$ and $x_{4}$, which leads to the
exact solutions in the neighborhoods of $x_{1}$ and $x_{4}$. The WKB
expansions are required to match the exact solution in $\left(  x_{-}%
,x_{2}-\delta\right)  $, $\left(  x_{3}+\delta,x_{+}\right)  $ and the
neighborhoods of $x_{1}$ and $x_{4}$. This matching strategy then gives a
family of quasinormal modes \cite{Karnakov2013},
\begin{align}
\left(  \omega_{bl}^{2}\right)  ^{R}  &  =V_{l}\left(  x_{b}\right)  +\left(
n+\frac{1}{2}\right)  \sqrt{2V_{l}^{(2)}\left(  x_{b}\right)  },\nonumber\\
\left(  \omega_{bl}^{2}\right)  ^{I}  &  =-\frac{\gamma_{n}^{2}}{2\int_{x_{2}%
}^{x_{3}}\frac{dx}{\sqrt{\left(  \omega_{bl}^{R}\right)  ^{2}-V_{l}\left(
x\right)  }}}\left(  e^{-2\int_{x_{1}}^{x_{2}}\sqrt{V_{l}\left(  x\right)
-\left(  \omega_{bl}^{R}\right)  ^{2}}dx}+e^{-2\int_{x_{3}}^{x_{4}}\sqrt
{V_{l}\left(  x\right)  -\left(  \omega_{bl}^{R}\right)  ^{2}}dx}\right)  ,
\label{eq:mode-b}%
\end{align}
where
\begin{equation}
\gamma_{n}=\pi^{1/4}\frac{1}{2^{n/2}\sqrt{n!}}\left(  \frac{2n+1}{e}\right)
^{\frac{1}{2}\left(  n+\frac{1}{2}\right)  }.
\end{equation}
Moreover, eqn. $\left(  \ref{eq:mode-b}\right)  $ indicates $\left(
\omega_{bl}^{2}\right)  ^{R}-V_{l}\left(  x_{b}\right)  $ $\propto n+\frac
{1}{2}$, which means that a large $n$ can drive the $\left(  \omega_{bl}%
^{2}\right)  ^{R}$ line away from the local minimum, making $x_{2}\in\left(
x_{-},x_{+}\right)  $ and $x_{3}\in\left(  x_{-},x_{+}\right)  $ impossible.
Therefore, eqn. $\left(  \ref{eq:mode-b}\right)  $ is only applicable to the
low-lying modes with small $n$. In FIG. \ref{Long-lived modes}, eqns. $\left(
\ref{eq:mode-b}\right)  $ is used to evaluate the lowest-lying modes $\left(
n=0\right)  $ of the hairy black hole with $\alpha=0.9$ and $Q=1.066$, which
are denoted by gray dots. Additionally, a direct integration numerically
solves eqn. $\left(  \ref{eq: Pert eq}\right)  $ for exact modes, which are
represented by blue dashed lines. We plot the real part $\omega_{bl}^{R}$
against $l$ in the left panel, which shows that the WKB results match well
with the numerical ones. As for the imaginary part $\omega_{b}^{I}$ in the
right panel, the WKB results differ a little from the numerical ones for a
small $l$. Nevertheless, the WKB results tend to approach the numerical ones
as $l$ increases, which demonstrates that the WKB approximation is accurate
enough when $l$ is large enough.

In the large $l$ limit, the real part $\omega_{bl}^{R}$ can be approximated
by,
\begin{equation}
\omega_{bl}^{R}\sim\Omega_{b}l-i\left(  n+\frac{1}{2}\right)  \lambda_{b}%
\sim\Omega_{b}l,\qquad l\gg1,
\end{equation}
where $\Omega_{b}$ is the angular velocity of the stable circular orbit at
$x=x_{b}$, and $\lambda_{b}$ is the corresponding Lyapunov exponent. Unlike
the global maximum case, the Lyapunov exponent $\lambda_{b}$ is purely
imaginary for the stable circular null geodesic, and hence contributes to the
real part of quasinormal modes. Nevertheless, the correction to $\omega
_{bl}^{R}$ due to $\lambda_{b}$ is negligible for a small value of $n$ in the
large $l$ limit. Hence, the real part of the low-lying quasinormal modes
trapped at the stable circular orbit is proportional to the corresponding
angular velocity. Schematically in the eikonal limit, the imaginary part
$\omega_{bl}^{I}$ can be expressed as
\begin{equation}
\omega_{bl}^{I}\sim-d\left(  e^{-c_{1}l}+e^{-c_{2}l}\right)  ,\qquad l\gg1,
\label{eq:reduced I of modes b}%
\end{equation}
where $d$, $c_{1}$ and $c_{2}$ are positive constants. It is observed that
$\omega_{bl}^{I}$ decays exponentially with respect to $l$ due to the double
potential barriers. In fact, $\omega_{bl}^{I}$ is related to the flux density
of leaking modes outside the double potential barriers. An exponentially small
value of $\omega_{bl}^{I}$ indicates that the double potential barriers trap
these modes in the potential valley with an exponentially large damping time.
The quasinormal modes living in the vicinity of the stable null geodesic are
thus dubbed as the long-lived modes. Since the long-lived modes can accumulate
around the stable null geodesic, their backreaction onto spacetime may render
the hairy black holes with a double-peak structure unstable
\cite{Cardoso:2014sna,Keir:2014oka,Guo:2021bcw}.

To study quasinormal modes at high excitation $\left(  n\gg1\right)  $, we
consider the $\left(  \omega_{bh}^{2}\right)  ^{R}$ line at some distance away
from the local minimum, which is illustrated in FIG. \ref{Veffx}. In the
vicinity of each turning point, the effective potential can be approximated by
a linear function. Then in the neighborhoods of the turning points, eqn.
$\left(  \ref{eq: Pert eq}\right)  $ can be exactly solved in terms of Airy
functions. Away from the turning points, WKB solutions provide a good
approximation. To obtain a complete solution, the WKB solutions should be
smoothly glued up by the exact solutions near the turning points, which leads
to the generalized Born-Sommerfeld quantization rule
\cite{Bender1999,Karnakov2013},
\begin{equation}
\int_{x_{2}}^{x_{3}}\sqrt{\omega_{bh}^{2}-V_{l}\left(  x\right)  }dx-\frac
{i}{4}\left(  e^{-2\int_{x_{1}}^{x_{2}}\sqrt{V_{l}\left(  x\right)
-\omega_{bh}^{2}}dx}+e^{-2\int_{x_{3}}^{x_{4}}\sqrt{V_{l}\left(  x\right)
-\omega_{bh}^{2}}dx}\right)  =\pi\left(  n+\frac{1}{2}\right)  .
\label{eq:BSq}%
\end{equation}
In the large $l$ limit, we extract the real part of quasinormal modes from the
quantization rule $\left(  \ref{eq:BSq}\right)  $, which reads
\begin{equation}
\int_{x_{2}}^{x_{3}}\sqrt{\left(  \omega_{bh}^{R}\right)  ^{2}-V_{l}\left(
x\right)  }dx=\pi\left(  n+\frac{1}{2}\right)  . \label{eq:real eq}%
\end{equation}
After the real part $\omega_{bh}^{R}$ is obtained, the imaginary part
$\omega_{bh}^{I}$ is then given by
\begin{equation}
\omega_{bh}^{I}=-\frac{1}{4\omega_{bh}^{R}\int_{x_{2}}^{x_{3}}\frac
{\sigma\left(  x\right)  }{\sqrt{\left(  \omega_{bh}^{R}\right)  ^{2}%
-V_{l}\left(  x\right)  }}dx}\left(  e^{-2\int_{x_{1}}^{x_{2}}\sqrt
{V_{l}\left(  x\right)  -\left(  \omega_{bh}^{R}\right)  ^{2}}dx}%
+e^{-2\int_{x_{3}}^{x_{4}}\sqrt{V_{l}\left(  x\right)  -\left(  \omega
_{bh}^{R}\right)  ^{2}}dx}\right)  , \label{eq:imag eq}%
\end{equation}
where
\begin{equation}
\sigma\left(  x\right)  =2\cos^{2}\left(  -\frac{\pi}{4}+\int_{x_{2}}^{x}%
\sqrt{\left(  \omega_{bh}^{R}\right)  ^{2}-V_{l}\left(  x\right)  }dx\right)
.
\end{equation}
Since the $\left(  \omega_{bh}^{2}\right)  ^{R}$ line is not close to the
local minimum, the left-hand side of eqn. $\left(  \ref{eq:real eq}\right)  $
becomes large in the eikonal limit, leading to a large $n$. Thus, eqns.
$\left(  \ref{eq:real eq}\right)  $ and $\left(  \ref{eq:imag eq}\right)  $
describe the quasinormal modes at high excitation with $n\gg1$. In this case,
$\sigma\left(  x\right)  $ oscillates dramatically between $x_{2}$ and $x_{3}%
$, and hence one has $\sigma\left(  x\right)  \approx1$ under the integration
of eqn. $\left(  \ref{eq:imag eq}\right)  $. Consequently, the imaginary part
of high excitation modes $\left(  \ref{eq:imag eq}\right)  $ can also be
schematically written as eqn. $\left(  \ref{eq:reduced I of modes b}\right)  $
in the eikonal limit. So for $l\gg1$, the quasinormal modes at high excitation
are also long-lived modes.

\subsection{Sub-long-lived Modes at Local Maximum}

\begin{figure}[ptb]
\begin{centering}
\includegraphics[scale=0.61]{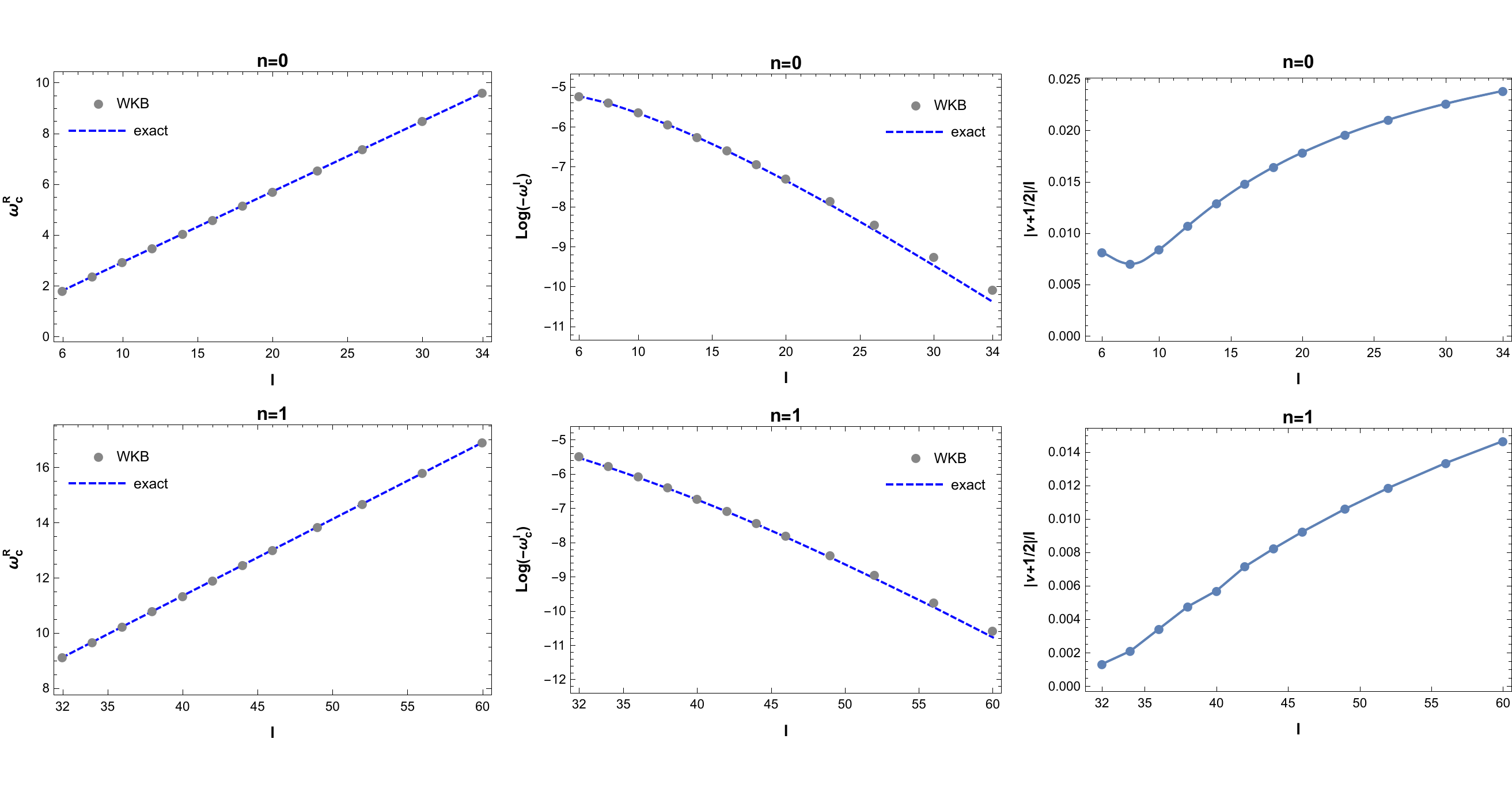}
\par\end{centering}
\caption{Real part (\textbf{Left}), imaginary part (\textbf{Middle}) and
$\left\vert \nu+1/2\right\vert /l$ (\textbf{Right}) of the quasinormal modes,
which are trapped at the local maximum of the effective potential, for the
branches of $n=0$ (\textbf{Upper}) and $n=1$ (\textbf{Lower}). The left column
shows that the real parts of the WKB results (gray dots) agree well with these
of exact modes (blue dashed lines). In the middle column, the imaginary parts
of WKB and exact results match well for small $l$, and start to deviate from
each other when $l$ becomes larger. As shown in the right column, the value of
$\left\vert \nu+1/2\right\vert /l$ grows as $l$ increases, which means that
the WKB approximation may not be accurate for a large $l$. }%
\label{LGmodes}%
\end{figure}

Finally, the quasinormal modes of frequency $\omega_{c}$, which live near the
smaller local maximum (i.e., the local maximum smaller than the global
maximum), are investigated. As illustrated in FIG. \ref{Veffx}, we consider
the $\left(  \omega_{c}^{2}\right)  ^{R}$ line lying in the vicinity of the
local maximum at $x=x_{c}$, which is associated with the outer unstable
circular null geodesic. Following the analogous strategy as before, the
effective potential is approximated with a parabola in the interval $\left(
x_{-},x_{+}\right)  $, and a linear function near the turning points $x_{1}$
and $x_{2}$. In addition, the turning points $x_{3}$ and $x_{4}$ are assumed
to lie in $\left(  x_{-},x_{+}\right)  $, which means that the WKB expansions
near $x_{3}$ and $x_{4}$ can match the exact solution with the parabolic
potential. Considering the boundary conditions and matching the WKB expansions
with the exact solutions near the turning points, we find the frequency of the
quasinormal modes is determined by (see Appendix \ref{sec:appd} for the
derivation)
\begin{equation}
\int_{x_{2}}^{x_{3}}\sqrt{\omega_{c}^{2}-V_{l}\left(  x\right)  }dx+\xi
+\frac{i}{4}e^{-2\int_{x_{1}}^{x_{2}}\sqrt{V_{l}\left(  x\right)  -\omega
_{c}^{2}}dx}=\left(  n+\frac{1}{4}\right)  \pi, \label{eq:QNMmatchingc}%
\end{equation}
where $\xi$ is defined by
\begin{equation}
e^{-2i\xi}=e^{-i\pi\nu+\left(  \nu+1/2\right)  }\left(  \nu+\frac{1}%
{2}\right)  ^{-\left(  \nu+\frac{1}{2}\right)  }\frac{\sqrt{2\pi}}%
{\Gamma\left(  -\nu\right)  },\nu+\frac{1}{2}=i\frac{\omega_{c}^{2}%
-V_{l}\left(  x_{c}\right)  }{\sqrt{-2V_{l}^{(2)}\left(  x_{c}\right)  }}.
\label{eq:xi}%
\end{equation}
Roughly speaking, the trapping in the potential valley and the tunneling
through the left potential barrier result in the first and last terms of eqn.
$\left(  \ref{eq:QNMmatchingc}\right)  $, respectively.

Since the distance between $x_{3}$ and $x_{4}$ is assumed to be small, the
value of $\nu+1/2$ should be tiny compared to $l$. In fact, the matching
condition $\left(  \ref{eq:QNMmatchingc}\right)  $ is valid to evaluate
quasinormal modes when the condition $\left\vert \nu+1/2\right\vert \ll l$ is
satisfied \cite{Karnakov2013}. Note that the valley of the effective potential
becomes deeper/shallower as $l$ increases/decreases. For a given $n$, a large
$l$ would drive the $\left(  \omega_{c}^{2}\right)  ^{R}$ line away from the
local maximum at $x=x_{c}$, thus making the first term in the left-hand side
of eqn. $\left(  \ref{eq:QNMmatchingc}\right)  $ small enough to satisfy eqn.
$\left(  \ref{eq:QNMmatchingc}\right)  $. However, if $\left(  \omega_{c}%
^{2}\right)  ^{R}$ is not close enough to the local maximum, the condition
$\left\vert \nu+1/2\right\vert \ll l$ can be violated, which indicates that
the WKB result $\left(  \ref{eq:QNMmatchingc}\right)  $ is not a good
approximation when $l$ is too large. In the case with large $l$, the effective
potential can be approximated by a linear function near the turning points
$x_{3}$ and $x_{4}$, which leads to the WKB results $\left(  \ref{eq:real eq}%
\right)  $ and $\left(  \ref{eq:imag eq}\right)  $. For an even larger $l$,
the $\left(  \omega_{c}^{2}\right)  ^{R}$ line can lie close to the minimum of
the potential, which corresponds to the aforementioned long-lived modes at low
excitations. On the other hand, when $l$ decreases, the first term in the
left-hand side of eqn. $\left(  \ref{eq:QNMmatchingc}\right)  $ requires that
the $\left(  \omega_{c}^{2}\right)  ^{R}$ line moves toward the local maximum
when $n$ is fixed. Interestingly, a too small $l$ can make $\left(  \omega
_{c}^{2}\right)  ^{R}$ greater than the local maximum, hence rendering the
turning points $x_{3}$ and $x_{4}$ unable to exist. So the WKB result $\left(
\ref{eq:QNMmatchingc}\right)  $ may cease to exist when $l$ is too small.

In FIG. \ref{LGmodes}, two branches of quasinormal modes are obtained using
eqns. $\left(  \ref{eq:QNMmatchingc}\right)  $ and $\left(  \ref{eq:xi}%
\right)  $, i.e., $n=0$ in the upper row and $n=1$ in the lower row. Moreover,
we also plot the value of $\left\vert \nu+1/2\right\vert /l$ as a function of
$l$ for each branch of the quasinormal modes in the right column, which checks
the validity of the WKB approximation. Compared with exact modes (blue dashed
lines), the real part of the quasinormal modes (gray dots) is well
approximated by the WKB method. By contrast, the imaginary part of the WKB
result matches that of exact modes well except when $l$ is too large, for
which, as displayed in the right column, the condition $\left\vert
\nu+1/2\right\vert \ll l$ is not well satisfied. It also shows that there
exists a lowest $l$ for each branch (e.g., $l=6$ for $n=0$ and $l=32$ for
$n=1$), below which the WKB results do not exist. Note that the exact modes of
$n=0$ presented in FIGs. \ref{Long-lived modes} and \ref{LGmodes} are
precisely the same. As expected, the exact modes are accurately described by
the WKB results $\left(  \ref{eq:QNMmatchingc}\right)  $ and $\left(
\ref{eq:mode-b}\right)  $ for small and large $l$, respectively.

\begin{table}[ptb]
\begin{centering}
\includegraphics[scale=0.48]{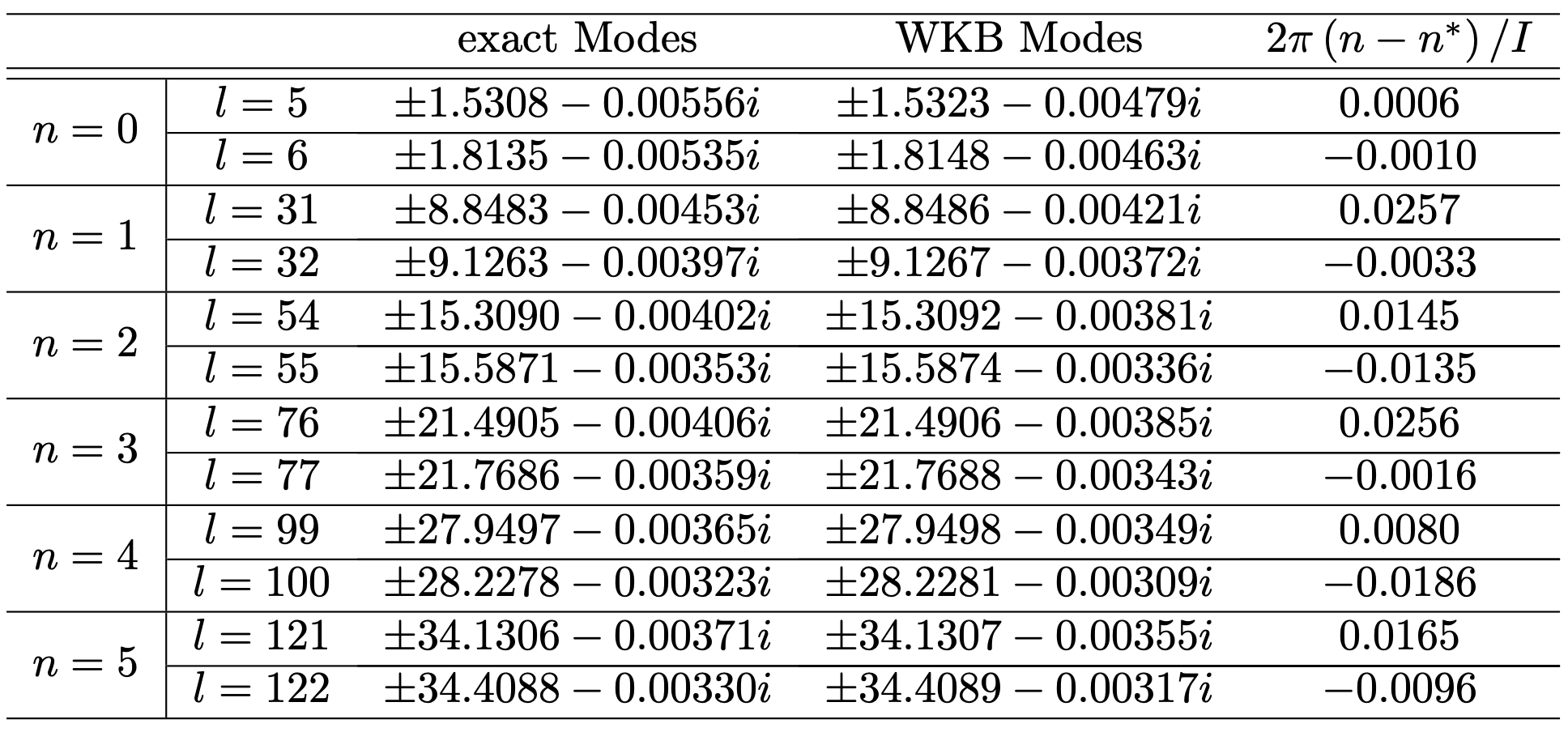}
\par\end{centering}
\caption{Pairs of sub-long-lived modes for different values of $n$. These
quasinormal modes are evaluated near the local maximum of the effective
potential, corresponding to an unstable circular null geodesic. The exact
modes are numerically obtained using a direct integration, which can be
approximated well by the WKB method for small magnitude of $2\pi\left(
n-n^{\ast}\right)  /I$.}%
\label{LPmodes}%
\end{table}

To find the relation between the quasinormal modes and the parameters of
circular null geodesics, we focus on the limit $\left\vert \nu+1/2\right\vert
\ll1$. In this limit, the matching condition $\left(  \ref{eq:QNMmatchingc}%
\right)  $ is further simplified (see Appendix \ref{sec:appd} for the
derivation),
\begin{align}
\left(  \omega_{c}^{2}\right)  ^{R}  &  \approx V_{l}\left(  x_{c}\right)
+\frac{\sqrt{-2V_{l}^{(2)}(x_{c})}2\pi I}{I^{2}+\pi^{2}/4}\left(  n-n^{\ast
}-\frac{e^{-2\int_{x_{1}}^{x_{2}}\sqrt{V_{l}\left(  x\right)  -\omega_{c}^{2}%
}dx}}{8I}\right)  ,\nonumber\\
\left(  \omega_{c}^{2}\right)  ^{I}  &  \approx-\frac{\log2}{2I}\sqrt
{-2V_{l}^{(2)}\left(  x_{c}\right)  }-\frac{\sqrt{-2V_{l}^{(2)}(x_{c})}\pi
^{2}}{I^{2}+\pi^{2}/4}\left(  n-n^{\ast}-\frac{Ie^{-2\int_{x_{1}}^{x_{2}}%
\sqrt{V_{l}\left(  x\right)  -\omega_{c}^{2}}dx}}{2\pi^{2}}\right)  ,
\label{eq:Re Im-c}%
\end{align}
where $I$ is
\begin{equation}
I=\log\left(  \sqrt{-2V_{l}^{(2)}\left(  x_{c}\right)  }\left(  x_{c}%
-x_{2}\right)  ^{2}\right)  +2\int_{x_{2}}^{x_{c}}\left(  \frac{\sqrt
{-2V_{l}^{(2)}\left(  x_{c}\right)  }}{2\sqrt{V_{l}\left(  x_{c}\right)
-V_{l}\left(  x\right)  }}-\frac{1}{\left(  x_{c}-x\right)  }\right)
dx+\left(  \gamma+\log2\pi\right)  , \label{eq:I}%
\end{equation}
with a constant $\gamma=0.5772$. Here, the number $n^{\ast}$ is defined as
\begin{equation}
n^{\ast}\equiv\frac{1}{\pi}\int_{x_{2}}^{x_{c}}\sqrt{V_{l}\left(
x_{c}\right)  -V_{l}\left(  x\right)  }dx+\frac{\log2}{8I}-\frac{1}{2},
\end{equation}
which can be interpreted as the number of a resonance filling the potential
well until $\left(  \omega_{c}^{2}\right)  ^{R}=V_{l}\left(  x_{c}\right)  $
in the eikonal limit. Moreover, the condition $|\nu+1/2|\ll1$ requires
\begin{equation}
\frac{2\pi|n-n^{\ast}|}{I}\ll1, \label{eq: delta n}%
\end{equation}
which provides a constraint on $n$ and $l$. For quasinormal modes of given
$n$, there appears to exist a pair of adjacent integers $l$ that well satisfy
the constraint $\left(  \ref{eq: delta n}\right)  $. For $n\leq5$, TABLE.
\ref{LPmodes} displays the quasinormal modes with such adjacent $l$. In this
case, it shows that the exact (obtained by a numerical direct integration
method) and WKB (obtained from eqn. $\left(  \ref{eq:Re Im-c}\right)  $)
results agree well with each other.

When $n$ is large enough, the value of $l$ satisfying the constraint $\left(
\ref{eq: delta n}\right)  $ can be arbitrarily large. In the large $l$ limit,
eqn. $\left(  \ref{eq:Re Im-c}\right)  $ reduces to
\begin{align}
\omega_{c}^{R}  &  \sim l\sqrt{V_{\text{geo}}\left(  x_{c}\right)  }%
=\Omega_{c}l,\nonumber\\
\omega_{c}^{I}  &  \sim-\frac{\log2}{2\log l}\lambda_{c}, \label{eq:QNMc}%
\end{align}
where we use $I\sim\log l$ for $l\gg1$. Here $\Omega_{c}$ is the angular
velocity of the unstable circular null geodesic at $x=x_{c}$, and $\lambda
_{c}$ is the corresponding Lyapunov exponent. For these quasinormal modes, the
turning points $x_{3}$ and $x_{4}$ are very close to $x_{c}$, and hence their
real part $\omega_{c}^{R}$ is proportional to the angular velocity $\Omega
_{c}$. Similar to the global maximum case, the Lyapunov exponent of the outer
unstable circular orbit contributes to the imaginary part of the quasinormal
modes. However, these quasinormal modes can temporarily trap in the potential
valley, which gives a logarithmically decaying factor $1/\log l$ in their
imaginary part. For this reason, this type of quasinormal modes is dubbed as
sub-long-lived modes.

\section{Conclusions}

\label{Sec:Conc}

In this paper, we studied quasinormal modes of a scalar field in hairy black
hole spacetime, where the scalar field is minimally coupled to the gravity
sector and non-minimally coupled to the electromagnetic field with an
exponential coupling function. Intriguingly, the hairy black holes have been
demonstrated to possess two unstable and one stable circular null geodesics on
the equatorial plane outside the event horizon, corresponding to two maxima
and one minimum of the geometric potential for null geodesic motion,
respectively. It showed that, apart from a constant prefactor, the effective
potential governing quasinormal modes of the scalar perturbation can be well
approximated by the geometric potential in the eikonal regime. To explore the
relation between quasinormal modes and the parameters of the circular null
geodesics, we used the WKB method to compute quasinormal modes living near the
global maximum, the smaller local maximum and the minimum of the effective potential.

In the large $l$ limit, the real part of these quasinormal modes was shown to
be proportional to the angular velocity of the corresponding circular null
geodesics, which implies that the quasinormal modes can be related to
particles traveling along the circular null geodesics. However, since the
imaginary part $\omega_{I}$ describes the perturbation decay timescale,
$\omega_{I}$ was found to have distinct behaviors, depending on the location
of the quasinormal modes. For the quasinormal modes near the global maximum,
$\omega_{I}$ can be interpreted as slowly leaking out of particles trapped at
the unstable circular null geodesics, and is related to the Lyapunov exponent,
which reflects the instability timescale of geodesic motion. On the other
hand, the effective potential valley between two maxima plays a key role in
determining the behavior of $\omega_{I}$ of quasinormal modes near the local
maximum and the minimum in the eikonal regime. When $l\gg1$, the depth of the
potential valley was found to be proportional to $l^{2}$. The quasinormal
modes living at the bottom of the potential valley (i.e., the minimum) are
metastable states with tunneling out through the high potential barriers,
which gives that $\omega_{I}$ decays exponentially with respect to $l$. There
appeared to be two contributions to $\omega_{I}$ of the quasinormal modes near
the local maximum, i.e., classical leaking out from the unstable circular null
geodesics, which makes $\omega_{I}$ proportional to the Lyapunov exponent, and
tunneling out through the global maximum barrier, which makes $\omega_{I}$
inversely proportional to $\log l$. Due to the exponential and logarithmic
suppressions in $\omega_{I}$, the quasinormal modes in the neighborhoods of
the minimum and local maximum can live for a long time, and hence were dubbed
as long-lived and sub-long-lived modes, respectively.

The long-lived modes may accumulate along the stable circular orbit, and
eventually develop a non-linear instability. Moreover for a spinning object,
the existence of long-lived modes may also trigger an ergoregion instability
at the linear level in the static limit \cite{Cardoso:2014sna}. These
instabilities imply that long-lived modes trapped at the stable circular orbit
could destabilize the background spacetime by their backreaction. In the
future studies, it is of great interest to further address the instabilities
of long-lived and sub-long-lived modes in a hairy black hole, and explore the
end point of a hairy black hole possessing three circular null geodesics in a
dynamic evolution.

\begin{acknowledgments}
We are grateful to Yiqian Chen and Qingyu Gan for useful discussions and
valuable comments. This work is supported in part by NSFC (Grant No. 12105191,
11947225 and 11875196). Houwen Wu is supported by the International Visiting
Program for Excellent Young Scholars of Sichuan University.
\end{acknowledgments}

\appendix

\section{Derivations of Eqns. $\left(  \ref{eq:QNMmatchingc}\right)  $ and
$\left(  \ref{eq:Re Im-c}\right)  $}

\label{sec:appd}

In this appendix, we follow \cite{Schutz:1985km,Bender1999,Karnakov2013} to
give derivations of eqns. $\left(  \ref{eq:QNMmatchingc}\right)  $ and
$\left(  \ref{eq:Re Im-c}\right)  $. As presented in FIG. \ref{Veffx}, we
consider the $\left(  \omega_{c}^{2}\right)  ^{R}$ line close to the local
maximum at $x=x_{c}$, and discuss the WKB solutions in different ranges of
$x$. When $x_{2}<x<x_{3}$, the WKB solution is
\begin{equation}
\Psi_{\text{I}}\left(  x\right)  \sim\left[  \omega_{c}^{2}-V_{l}\left(
x\right)  \right]  ^{-1/4}\sin\left(  \int_{x_{2}}^{x}\sqrt{\omega_{c}%
^{2}-V_{l}\left(  x^{\prime}\right)  }dx^{\prime}+\frac{\pi}{4}+\frac{i}%
{4}e^{-2\int_{x_{1}}^{x_{2}}\sqrt{V_{l}\left(  x\right)  -\omega_{c}^{2}}%
dx}\right)  . \label{eq:Psi1}%
\end{equation}
For $x>x_{4}$, the outgoing WKB solution is
\begin{equation}
\Psi_{\text{III}}\left(  x\right)  \sim\left[  \omega_{c}^{2}-V_{l}\left(
x\right)  \right]  ^{-1/4}e^{i\int_{x_{4}}^{x}\sqrt{\omega_{c}^{2}%
-V_{l}\left(  x^{\prime}\right)  }dx^{\prime}}. \label{eq:Psi3}%
\end{equation}
In the interval $\left(  x_{-},x_{+}\right)  $, the effective potential is
approximated by a parabolic expansion, for which the perturbative equation
$\left(  \ref{eq: Pert eq}\right)  $ can be exactly solved. To match the
WKB\ solution $\left(  \ref{eq:Psi3}\right)  $ in the region $\left(
x_{4},x_{+}\right)  $, the exact solution $\Psi_{\text{II}}\left(  x\right)  $
is then given by
\begin{equation}
\Psi_{\text{II}}\left(  x\right)  \sim D_{\nu}\left(  t\right)  ,
\label{eq:Psi2}%
\end{equation}
where $D_{\nu}\left(  t\right)  $ represents the parabolic cylinder function,
$\nu$ is defined in eqn. $\left(  \ref{eq:xi}\right)  $, and $t\equiv
e^{i\pi/4}[-2V_{l}^{(2)}(x_{c})]^{1/4}(x-x_{c})$. On the other hand, to match
the solution $\left(  \ref{eq:Psi2}\right)  $ with the exact solution
$\Psi_{\text{II}}\left(  x\right)  $ in the region $\left(  x_{-}%
,x_{3}\right)  $, the WKB solution for $x_{2}<x<x_{3}$ should be
\begin{equation}
\Psi_{\text{I}}^{\prime}\left(  x\right)  \sim\left[  \omega_{c}^{2}%
-V_{l}\left(  x\right)  \right]  ^{-1/4}\sin\left(  \int_{x}^{x_{3}}%
\sqrt{\omega_{c}^{2}-V_{l}\left(  x^{\prime}\right)  }dx^{\prime}+\xi
+\frac{\pi}{2}\right)  , \label{eq:Psi11}%
\end{equation}
where $\xi$ is defined in eqn. $\left(  \ref{eq:xi}\right)  $. To satisfy
$\Psi_{\text{I}}\left(  x\right)  \propto\Psi_{\text{I}}^{\prime}\left(
x\right)  $ for $x\in(x_{2},x_{3})$, the sum of the components of sine
functions in eqns. $\left(  \ref{eq:Psi1}\right)  $ and $\left(
\ref{eq:Psi11}\right)  $ must be a multiple of $\pi$, which gives eqn.
$\left(  \ref{eq:QNMmatchingc}\right)  $.

In the limit $|\nu+1/2|\ll1$, the first term of the left-hand side of eqn.
$\left(  \ref{eq:QNMmatchingc}\right)  $ becomes
\begin{align}
\int_{x_{2}}^{x_{3}}\sqrt{\omega_{c}^{2}-V_{l}\left(  x\right)  }dx  &
=\int_{x_{2}}^{x_{-}}\sqrt{\omega_{c}^{2}-V_{l}\left(  x\right)  }%
dx+\int_{x_{-}}^{x_{3}}\sqrt{\omega_{c}^{2}-V_{l}\left(  x\right)
}dx\nonumber\\
&  \approx\int_{x_{2}}^{x_{c}}\sqrt{V_{l}(x_{c})-V_{l}(x)}dx+\frac{ia}%
{2}\left(  \log\left(  ia\right)  -1-\log\left[  \sqrt{-2V_{l}^{(2)}(x_{c}%
)}\left(  x_{c}-x_{2}\right)  ^{2}\right]  \right) \nonumber\\
&  -ia\int_{x_{2}}^{x_{c}}\left(  \frac{\sqrt{-2V_{l}^{(2)}(x_{c})}}%
{2\sqrt{V_{l}(x_{c})-V_{l}(x)}}-\frac{1}{\left(  x_{c}-x\right)  }\right)  dx.
\label{eq:int1}%
\end{align}
where we replace the effective potential by a parabola in the region $\left(
x_{-},x_{c}\right)  $, and $a=\nu+1/2$. Moreover, $\xi$ in eqn. $\left(
\ref{eq:QNMmatchingc}\right)  $ can be expanded as
\begin{equation}
\xi\approx\frac{i}{2}\left(  -i\pi a-a\ln a+\frac{\ln2}{2}+\frac{i\pi}%
{2}+a\left[  1-\left(  \gamma+\ln2+\ln\pi\right)  \right]  \right)  ,
\label{eq:expr of xi}%
\end{equation}
where $\gamma=0.5772$. From eqns. $\left(  \ref{eq:QNMmatchingc}\right)  $,
$\left(  \ref{eq:int1}\right)  $ and $\left(  \ref{eq:expr of xi}\right)  $,
one can extract the real and imaginary parts of $\omega_{c}^{2}$, which gives
eqn. $\left(  \ref{eq:Re Im-c}\right)  $.

\bibliographystyle{unsrturl}
\bibliography{ref}

\end{document}